\documentstyle[epsfig]{aipproc}

\begin{document}
\begin{flushright}
ADP-00-20/T403
\end{flushright}

\title{What Can We Learn from QED at Large Couplings?}

\author{A. W. Schreiber$^a$, R. Rosenfelder$^b$ and C. Alexandrou$^c$}

\address{$^a$Department of Physics and Mathematical Physics and
           Research \\Centre for the Subatomic Structure of Matter\\
           University of Adelaide, Adelaide, S. A. 5005, Australia}
\address{$^b$Paul Scherrer Institut, CH-5232 Villigen PSI, 
         Switzerland}
\address{$^c$Department of Physics, University of Cyprus\\
         CY-1678  Nicosia, Cyprus}

\maketitle

\begin{abstract}
In order to understand QCD at the energies relevant to hadronic
physics one requires analytical methods for dealing with relativistic
gauge field theories at large couplings.  Strongly coupled quenched
QED provides an ideal laboratory for the development of such
techniques, in particular as many calculations suggest that -- like
QCD  -- this theory has a phase with broken chiral symmetry.  In this talk
we report on a nonperturbative variational calculation of the
electron propagator within quenched QED and compare results to
those obtained in other approaches. We find  surprising differences 
among these results.

\end{abstract}

\section*{Introduction}

It is well known that the content of a relativistic field 
theory (let us take QCD as an example) may be expressed in 
terms of functional averages
of operators of the type
\begin{equation}
\int {\cal D}[\overline \Psi, \Psi, A] \> 
{\cal O}[\overline \Psi, \Psi, A] \>
e^{-S_{\rm QCD}[\overline \Psi, \Psi, A]}\;\;\;.
\label{eq: aws path integral}
\end{equation}
Even though it is not possible to perform these integrals exactly, we
have learnt an awful lot over the years by studying Eq.~(\ref{eq: aws
path integral}) in various approximations.  For example, if there is a
small parameter (e.g. $g$, $1/N_c$\ldots) one may develop well-ordered
expansions of Eq.~(\ref{eq: aws path integral}).  Alternatively one  can use stationary
phase methods to approximate the integral --  relevant, for example, if
one is interested in 
elucidating the importance of classical configurations.
Also, these methods are used in order to gain understanding of the behaviour of
very high orders of perturbation theory (HOPT).  One may study the
symmetries of the theory (e.g. chiral perturbation theory) in order to relate
observables, or one can study its equations of motion
(e.g. Dyson-Schwinger equation (DSE) studies).  Finally, much progress
has been made in recent years by actually evaluating
Eq.~(\ref{eq: aws path integral}) directly by discretizing it, as is done in
Lattice QCD.  

We have reported elsewhere on yet another technique, the so-called
``worldline variational approach'', in which the path integral is
approximated, in a rigorous and systematically correctable way, via a
variational principle.  This approach was initially applied to
relativistic field theory in a scalar model~\cite{aws wc} and more
recently to quenched QED~\cite{aws qed2}.   The common property that
QED  shares with QCD 
is that they are both renormalizable gauge field theories, 
in contrast to the scalar
model which is super-renormalizable.    
Of course QCD, even after quenching (i.e.
neglecting pair creation), differs from quenched QED 
by the fact that it is asymptotically free whereas
quenched QED has a
constant coupling.
 However,  many of the above-mentioned approximative schemes for
dealing with Eq.~(\ref{eq: aws path integral}) should be applicable
within either theory.  Quenched QED therefore serves, and is often
used, as a test-ground for these methods.  In this talk we compare results
for a particular quantity characterising quenched QED, namely its
anomalous mass dimension, obtained via a) the variational approach, b)
via perturbation theory, c) via Dyson-Schwinger equation studies and
d) via HOPT.  Surprisingly, we find that these results are not
entirely consistent with each other, which indicates that our present
understanding of these methods may be incomplete.

\begin{figure}[htb]
\begin{center}
\epsfig{angle=90,height=8cm,file=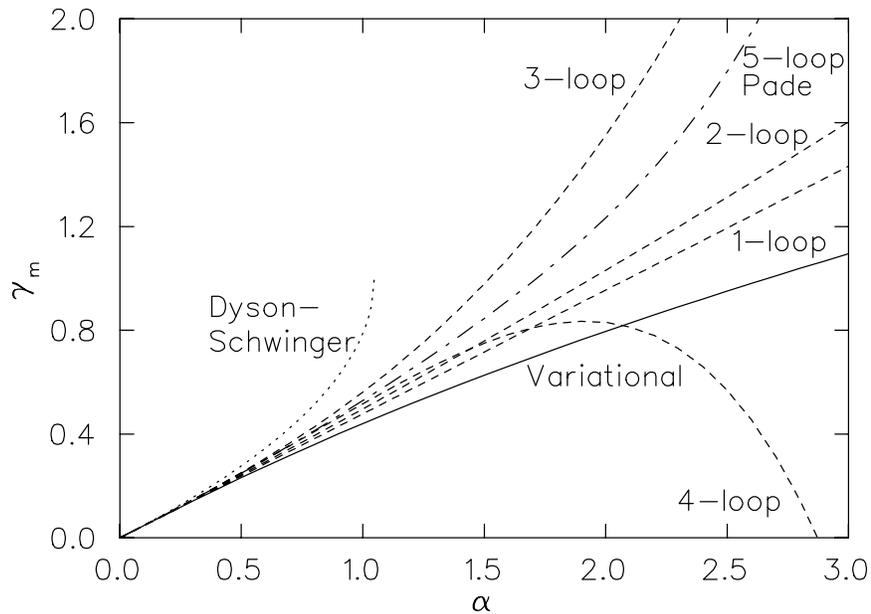}
\end{center}
\caption{Anomalous mass dimension $\gamma_m$ as function of the
coupling constant $\alpha$ in quenched QED. The variational result
(Eq.~\ref{eq: aws gamvar}) is shown as a solid curve.  The curves labeled ``n-loop''
show the result up to n-loop perturbation theory.  The Pad\'e
estimation of the 5-loop result is shown as a dot-dashed line and finally, the
solution from the Dyson-Schwinger equations in rainbow approximation
is indicated as a dotted curve.}
\label{fig: aws mass dim}
\end{figure}

In Ref.~\cite{aws qed2} we derived from the variational approach an analytic,
implicit, result for the anomalous mass dimension $\gamma_m(\alpha)$
of quenched, dimensionally regularized, QED within the MS scheme:
\begin{equation}
\alpha \>=\> {4 \over 3}\> (1 + \gamma_m^{\rm var}) \>\cot 
{\pi/2 \over 1 + \gamma_m^{\rm var}}\;\;.
\label{eq: aws gamvar}
\end{equation} 
This is a remarkable result.  At small couplings it should
(approximately) agree with the known result of perturbation
theory~\footnote{$\gamma_m(\alpha)$ has been calculated to fourth
order in $\alpha$ within SU(N)~\cite{aws alpha4}; by suitable choice of Casimirs
(i.e. $C_A=0$, $C_F=1$ and, for a quenched theory, $N_f=0$) the result
for U(1) may be extracted from this.  Recently, Broadhurst~\cite{aws broadhurst}
has also calculated $\gamma_m(\alpha)$ to fourth order directly within QED}.
Indeed, as may be seen in Fig.~\ref{fig: aws mass dim} (taken from
Ref.~\cite{aws qed2}), the variational result agrees with perturbation
theory to within $\approx$ 20 \% for couplings where 4$th$ order
perturbation theory appears applicable, i.e. for $\alpha$ less than
about 1. We have also plotted an estimate of perturbation theory up to 5$th$ order
term of the perturbative expansion~\footnote{The 5$th$ order  estimate has
been derived using the Pad\'e approximation methods
of Ref.~\cite{aws Pade}}.

It is interesting to note that for a value of $\alpha$ around $1$ the various
orders of perturbation theory appear to `fan out'.  It is tempting to
speculate that this effect arises because of a finite radius
of convergence of the perturbation expansion.  This would be unexpected:
the general wisdom from HOPT studies is that, as a first rough estimate, the
$n^{th}$ order contribution to a perturbative expansion should scale roughly
like the number of diagrams at that order.  For the quenched QED propagator,
the number of diagrams at order $n$ is given by 
\begin{equation}
N_n \> = \> { (2 n)! \over 2^n \> n!} \> = \> (2 n - 1)!! 
 \> \stackrel{n \to \infty}{\longrightarrow} \>
2^{n+1/2}\> e^{-n} \> n^n
\label{eq: aws factorial growth}
\end{equation}
and hence the radius of convergence (= 
$\lim_{n \rightarrow \infty} |N_n|^{-1/n}$) of the perturbative expansion
vanishes.  More sophisticated analyses\cite{aws large orders} confirm
this general result (although, apparently, the high order behaviour of
$\gamma_m$ does not itself appear to ever have been calculated
explicitly within quenched QED).

The final curve plotted in Fig.~\ref{fig: aws mass dim} is the
prediction of a Dyson-Schwinger equation calculation within the
rainbow approximation.  This was obtained, for the dimensionally
regularized theory within the MS scheme, in Ref.~\cite{aws qed2,aws
qed3} and was found to be identical with the well known result for
$\gamma_m$ in the theory regularized with a hard cut-off, i.e.
$\gamma_m^{\rm DS} \> = \> 1 \> - \> \sqrt{1 - {3 \over \pi} \alpha
}$.  At small couplings this agrees with perturbation theory but then
diverges from the latter in a region where (4$^{th}$ order)
perturbation theory still appears to be applicable.  Above $\alpha \>
=\> \pi/3$, $\gamma_m^{\rm DS}$ becomes complex, which is also the
value $\alpha_{cr}$ of the coupling at which the famous chiral
symmetry breaking of quenched QED occurs.  Note also that the
perturbative expansion of $\gamma_m^{\rm DS}$ has a finite radius of
convergence given by $\alpha_{con}^{\rm DS} \> = \> \alpha_{cr}$.  It
is, however, not a great surprise that $\alpha_{con}^{\rm DS}$ is
finite in this calculation as the rainbow approximation contains
exactly one diagram at each order and hence does not reproduce the
factorial growth (Eq. ~\ref{eq: aws factorial growth}) in the number
of diagrams.  Nevertheless, the coincidence of $\alpha_{con}^{\rm DS}$
and $\alpha_{cr}^{\rm DS}$ is interesting and one wonders if it will
persist in Dyson-Schwinger calculations when going beyond the rainbow
approximation.  Also, if in the exact theory $\alpha_{con}$ actually
vanishes one wonders what will happen to $\alpha_{cr}$ in that case.

The variational result plotted in Fig.~\ref{fig: aws mass dim}
appears to show no sign of chiral symmetry breaking. Indeed it is easy
to derive that at large couplings $\gamma_m^{\rm var} \rightarrow \sqrt{3 \pi/8}
\>\sqrt{\alpha}$.  However, it turns out that a perturbative expansion
of $\gamma_m^{\rm var}$ also has a finite radius of convergence, not too different
from the one mentioned above.  We shall discuss this, as well as the large order 
behaviour of  the perturbative expansion of $\gamma_m^{\rm var}$, below.  It is
important to note that in this case a finite radius of convergence is not
a trivial result of the approximation, as it was for rainbow QED:  it can be 
shown~\cite{aws wc} that each diagram, at any order in perturbation theory,
is represented in some approximate way within the variational approximation; i.e.
a perturbative expansion of $\gamma_m^{\rm var}$ would receive contributions from
$(2 n - 1)!!$ diagrams at order $n$.

\section*{The radius of convergence of $\gamma_m^{\rm var}$}

\begin{figure}[htb]
\begin{center}
\setlength {\unitlength} {1mm}
\begin{picture}(70,70)(15,00)
\epsfig{height=7cm,file=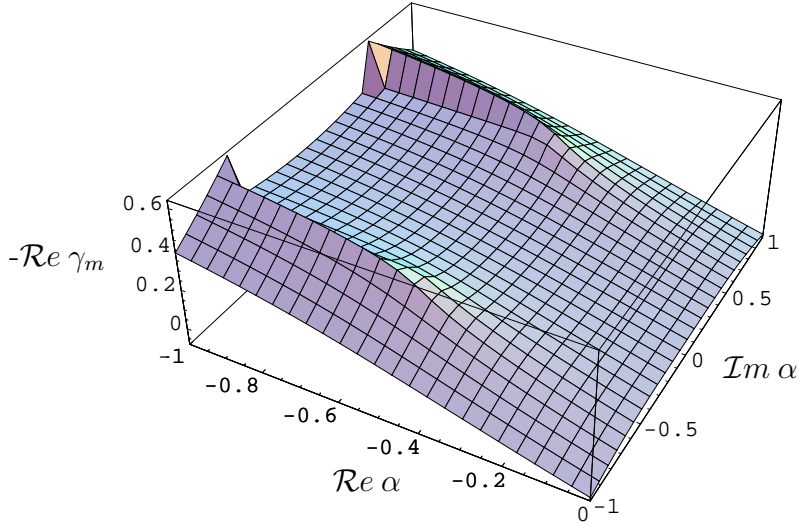}
\put(-60,5){${\cal R}e \> \alpha$}
\put(-8,20){${\cal I}m \> \alpha$}
\put(-103,35){-${\cal R}e \> \gamma_m$}
\end{picture}
\end{center}
\caption{Solution of Eq.~(\protect\ref{eq: aws gamvar}) for complex $\alpha$.  The figure
was obtained by using Newton's method, with an initial seed value of 
$\gamma_m=0$.  No interesting analytic structure is found at positive
${\cal R}e \>\alpha$, hence this region is not shown.}
\label{fig: aws branch cut}
\end{figure}

Equation~(\ref{eq: aws gamvar}) is an implicit equation for $\gamma_m^{\rm var}$
which can easily be solved numerically for arbitrary $\alpha$.  Importantly
it can also be solved for complex $\alpha$, so the determination of $\alpha_{con}$
amounts to a search for nontrivial analytic structure in the complex $\alpha$
plane.  In Fig.~\ref{fig: aws branch cut} we show a plot of the (negative) real part
of $\gamma_m$ as a function of the  (complex) coupling.  Branch cuts, limiting the
region of convergence of the perturbation expansion of  $\gamma_m^{\rm var}$, 
are clearly 
visible for negative ${\cal R}e \> \alpha$.  Note that, as opposed to the 
Dyson-Schwinger equation result $\gamma_m^{\rm DS}$, there are no cuts on the real
axis (also for positive ${\cal R}e \> \alpha$, which is not shown in 
Fig.~\ref{fig: aws branch cut}). Hence, in 
Fig.~\ref{fig: aws mass dim}, $\gamma_m^{\rm var}$ is real for all $\alpha$
while $\gamma_m^{\rm DS}$ terminates at $\alpha_{con}$~\cite{aws mrs}.

The position of the cuts in Fig.~\ref{fig: aws branch cut} can be obtained
straightforwardly by searching for the value of $\gamma_m$ at which 
Eq.~(\ref{eq: aws gamvar}) has two distinct solutions infinitesimally close
to each other; i.e. the value of $\gamma_m^{\rm var}$ at which the derivative with respect
to $\gamma_m^{\rm var}$ of Eq.~(\ref{eq: aws gamvar}) vanishes.  One obtains that
the branchpoints are located at $\alpha \> = \> -0.496127 \pm 0.619172 \> i$, which
yields a radius of convergence for the perturbation expansion of $\gamma_m^{\rm var}$
of $\alpha_{con}^{\rm var} \> = \> 0.79342$.

\section*{The behaviour of $\gamma_m^{\rm var}$ at large orders in
perturbation theory}

The radius of convergence of the  perturbative expansion of the anomalous 
mass dimension can also be obtained, of course, by directly examining the behaviour
of its expansion coefficients at large orders, i.e. if we expand
\begin{equation}
\gamma_m \> = \> \sum_{n=1}^{\infty} c_n \> \alpha^n\;\;\;,
\label{eq: aws perturbative expansion}
\end{equation}
then radius of convergence is given by
\begin{equation}
\alpha_{con} \> = \> \lim_{n \rightarrow \infty}\> |c_n|^{-1/n}\;\;\;.
\end{equation}
For $n$ up to $\approx 30$, it is possible to obtain these expansion
coefficients by direct substitution of Eq.~(\ref{eq: aws perturbative expansion})
into Eq.~(\ref{eq: aws gamvar}).  The results are tabulated in 
Table~\ref{table: aws cn}, where it is clearly seen that although $|c_n|^{-1/n}$
does perhaps tend to a finite limit, the rate of convergence is rather slow
(and non-uniform).

\begin{center}
\begin{table}[b]
\caption{Expansion coefficients of the perturbative expansion of 
$\gamma_m^{\rm var}$.}
\label{table: aws cn}
\begin{tabular}{ccccccc}
n & 5 & 10 & 15 & 20 & 25 & 30\\\\
$ |c_n|^{-1/n}$ & 2.00 & 1.31& 1.16& 1.11& 1.02& 1.16\\
\end{tabular}
\end{table}
\end{center}

\begin{figure}[htb]
\begin{center}
\setlength {\unitlength} {1mm}
\begin{picture}(50,50)(-65,00)
\epsfig{height=4cm,file=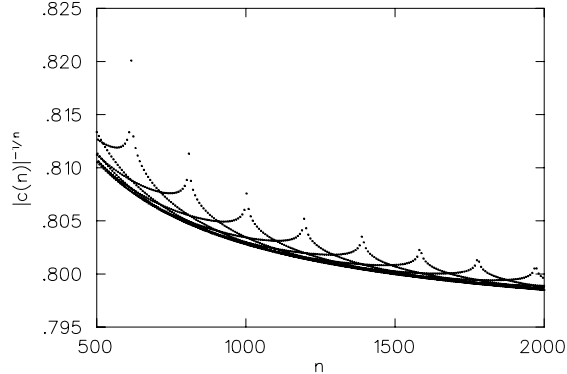}
\end{picture}
\end{center}
\caption{The coefficients $c_n$ of the perturbative expansion of
$\gamma_m^{\rm var}$. }
\label{fig: aws cn}
\end{figure}

In order to find the $c_n$'s for higher values of $n$, it proves to be 
advantageous to convert Eq.~(\ref{eq: aws gamvar}) into a differential
equation in order to eliminate the cotangent.  We
obtain
\begin{equation}
1 + \gamma_m^{\rm var} \> = \> \left ( \alpha + {2 \pi \over 3} \right )
{\gamma_m^{\rm var\>}} ' -
{3 \pi \over 8} \alpha^2  \left ({1  \over 1 + \gamma_m^{\rm var}} \right )'\;\;\;.
\end{equation}
If we now substitute Eq.~(\ref{eq: aws perturbative expansion})
for $\gamma_m^{\rm var}$, and $\sum_{n=0}^{\infty} a_n \> \alpha^n$ for 
$1/(1+\gamma_m^{\rm var})$ (hence  $a_0 \> = \> 1$ and  
$a_n \> = \> - \sum_{k=1}^{n-1} \> c_{n-k} \> a_n$)
we may solve for the coefficients $c_n$ and $a_n$ in an iterative manner.
The results are shown in Fig.~\ref{fig: aws cn}.

The slow rate of convergence, as well as its non-uniformity, is clearly
visible in this figure.  Indeed, numerically one finds that for large $n$ 
the points in Fig.~\ref{fig: aws cn} are fitted exceedingly well 
by the functional form
\begin{equation}
c_n \> \approx \> \left (\alpha_{con}^{\rm var}\right )^{-n} {e^{-\beta} \over n^{3/2}} 
\sin \left [ \left (a + {5 \pi \over 7}\right ) n \> - \> {3 \pi \over 7} 
  \> + \>b \right ]\;\;\;, 
\end{equation}
with $\beta \approx 1.376$, $a \approx 2.32 \times 10^{-3}$ and
$b \approx -8.268 \times  10^{-2}$.  These values of $c_n$ may be
compared to the equivalent expansion coefficients obtained from
the Dyson-Schwinger equation result:
\begin{equation}
c_n^{\rm DS} \> \approx \> \left (\alpha_{con}^{\rm DS}\right )^{-n}{e^{-1.27} \over n^{3/2}} \;\;\;.
\end{equation}
In fact, the variational result and the rainbow Dyson-Schwinger result 
for the $c_n$'s are surprisingly similar, not only in their functional form
but even in the numerical coefficients.  The main difference is the occurance
of the sine function in the former.  It is because of this sine that the
branchcut, which for $\gamma_m^{\rm DS}$ lies on the real axis, has moved into
the complex plane for $\gamma_m^{\rm var}$.  

\section*{Conclusion}

In this contribution we have compared predictions for the anomalous
mass dimension of quenched QED obtained through the use of a variety of techniques:
the worldline variational approach, perturbation theory up to
$O[\alpha^4]$ (as well as a Pad\'e estimate for the $O[\alpha^5]$
term), rainbow Dyson-Schwinger equation studies and general
expectations from studies of high order perturbation theory.  Both the
rainbow DSE's and the variational approach yield a cut in $\gamma_m$
as a function of the coupling.  In the former this cut is on the real
axis, and is associated with chiral symmetry breaking, while in the
latter it has moved (a long way) off the real axis, so that in that
calculation there is no obvious sign of chiral symmetry breaking.  In
either case, the cuts necessitate a finite radius of convergence for
the perturbative expansion of $\gamma_m(\alpha)$, and one can argue
that circumstantial evidence for this may be seen in the perturbative
result as well.  General expectations from HOPT, on the other hand,
suggest that the radius of convergence should be zero.  Clearly, as
quenched QED is the prototype gauge theory for the investigation of
chiral symmetry breaking, efforts should be made to clarify these
disagreements.  It may even be the case that the eventual resolution
of these issues will teach us something about the limitations of
one or more of these approximate techniques for dealing with nonperturbative gauge
field theories.

\def\Journal#1#2#3#4{{#1} {\bf #2}, #3 (#4)}

\def\NCA{\em Nuovo Cimento}
\def\NIM{\em Nucl. Instrum. Methods}
\def\NIMA{{\em Nucl. Instrum. Methods} A}
\def\NPA{{\em Nucl. Phys.} A}
\def\NPB{{\em Nucl. Phys.} B}
\def\PLB{{\em Phys. Lett.}  B}
\def\PRL{\em Phys. Rev. Lett.}
\def\PRD{{\em Phys. Rev.} D}
\def\PRA{{\em Phys. Rev.} A}
\def\PR{{\em Phys. Rev.}}
\def\ZPC{{\em Z. Phys.} C}
\def\PREP{{\em Phys. Rep.}}
\def\IJMPE{{\em Int. Journ. Mod. Phys.} E}
\def\JETP{{\em Sov. Phys. JETP}}

\end{document}